\documentstyle[aps]{revtex}
\begin{document}

\draft

\title{Reconstruction of a first-order phase transition from computer
simulations of individual phases and subphases}

\author{Martin Ebeling and Walter Nadler}

\address{Institut f\"ur Theoretische Chemie, Universit\"at
T\"ubingen, Auf der Morgenstelle 8, D-72076 T\"ubingen, Germany}

\maketitle

\begin{abstract}

We present a new method for investigating first-order
phase transitions using Monte Carlo simulations. It relies on the
multiple-histogram method and uses solely histograms of individual
phases.
In addition, we extend the method to include histograms of subphases.

 The free energy difference between phases, necessary for attributing
the correct statistical weights to the histograms, is determined by a
detour in
control parameter space via auxiliary systems with short relaxation
times. We
apply this method to a recently introduced model for structure
formation in
polypeptides for which other methods fail.
\end{abstract}

\pacs{02.70.Lq, 02.50.Ng, 64.60.-i, 87.15.By}

\narrowtext

At first-order phase transitions, Monte Carlo (MC) simulations
encounter a
particular problem of critical {slowing} down: since transitions
between the
co-existing phases are very rare, the relaxation time is extremely
long and
usually increases exponentially with system size. In order to obtain
equilibrated data via MC simulations for a  system of appreciable
size near the
transition, prohibitively long simulations would have to be
performed.

A convenient way to visualize the problem as well as to analyze
simulation
results is to employ histograms. Since the Hamiltonian and
most observables of interest are functions of one or a few order
parameters,
such as magnetization or internal energy, a histogram of the
frequency of their
occurence, recorded during the simulation, is sufficient to generate
all
information of interest.  Using the single- and the
multiple-histogram method (SHM/MHM) \cite{FS}, one or several
histograms
determined for a system at specific sets of control parameters can be
used
efficiently to predict the system behaviour over a wide range of
control
parameter values. In such histograms the co-existence
of phases corresponds to several peaks, see {\it e.g.} Fig.\
\ref{hist}.
The infrequent switching between individual
phases makes it difficult to generate
a single equilibrated histogram covering all phases, {\it i.e.}
showing all peaks
correctly.

Most approaches to this problem  introduce suitable changes to the
system
Hamiltonian so that transition
{states}, which are encountered only rarely in the original system,
will be sampled much more frequently in a simulation, thereby leading
also to a
higher frequency of switching between
phases\cite{BN,Lee93,KW93,MP92}.
Common to all these methods is the idea that in a {\it single}
simulation a
histogram can be determined which covers all interesting regions of
state space.
A drawback is that possibly a plethora of parameters, which render
the
necessary changes in the system Hamiltonian, have to be optimized.

In this contribution we want to present a new approach to the
problem. It is
based on the realization that an {equilibrated} histogram of a system
confined to an
individual phase, {\it i.e.} an individual peak in the full
histogram,
can be generated much more easily.
Under those conditions the
relaxation times are usually small, at least much smaller than the
switching
times between phases in the critical regime.
For a first-order phase transition, these single-phase histograms
already cover
all  relevant system states because, even at the critical point,
transition states between the phases are encountered only very
infrequently,
and, hence, their contribution to the partition function is
negligible.
Several single-phase histograms can, therefore, be combined by the
MHM to reconstruct the
 complete histograms at or near the first-order transition.

However, there is a technical problem involved: the relative
contributions of various  histograms, in effect the free energy
differences
between them,
have to be determined. This is a notoriously difficult task. A
prerequisite for
their correct determination, $e.g.$ by the acceptance
ratio method (ARM) developed by Bennett in his definitive treatment
of the
subject\cite{Bennett76}, is that the histograms in question overlap
at least
partially. This is usually not the case, particularly not for systems
with strong first-order
transitions. To overcome this difficulty, we use a detour in
control parameter space via auxiliary systems  with much shorter
relaxation times, which resemble the actual system to an extent
sufficient to
guarantee significant overlaps between actual and auxiliary
histograms\cite{RHRitt}.

We will apply this scheme to the  temperature-driven first-order
phase
transition in a recently introduced simple model  of secondary and
tertiary
structure formation in polypeptides\cite{EN}, where the
methods of Ref.\ \cite{BN,Lee93,KW93,MP92} fail.

 The conformation of a
polypeptide of length $L$ is represented by a string of local
conformations
$\sigma_{i} = h$, $c^+$, or $c^0$, $i = 1, \ldots, L$. The local
conformation
$h$ corresponds to residues with dihedral angles characteristic of
$\alpha$
helices. Any three successive monomers in helical conformation are
spanned by a
hydrogen bond with energy $E_{HB}<0$. A string of $l$ consecutive
$h$-residues ($l
\geq 3$) forms a helix of length $l$. The $c^0$ and $c^+$ residues
denote random
coil conformations. Two helices separated solely by
$c^0$ residues are taken to interact with each other, whereas helices
with at least one
residue with conformation other than $c^0$ between them are not.
The number of contacts between two interacting helices, which
determines the
interaction energy, is taken to be equal to the length of the shorter
helix.
We set the interaction energy parameter to $k=0.6\, E_{HB}$. Since
the conformation space volume $\Omega(h)$ accessible to $h$ residues
is smaller than that for non-$h$ residues, the conversion $c^{+} \,
\leftrightarrow \, h$, for example, is accompanied by a change in
conformational entropy. We define $\Delta S(\sigma_{i}) =
k_{B}\ln\lbrack \Omega(\sigma_{i})/\Omega(c^{+})\rbrack$, and assume
 equal
statistical weights for the two random coil conformations, {\it
i.e.},
$\Delta S(c^0) = \Delta S(c^+) = 0$. The helix-coil
transition in experimentally studied homo(poly)amino acids is
adequately
described by a value of $\Delta S(h)/k_{B} = -4.26 + \ln(2) \approx
-3.57$,
derived from experimental data
 first discussed by Zimm and Bragg\cite{ZB59}.

Using the order parameter vector of a chain conformation
$\{\sigma_{i}\}$,  ${\underline S} (\{\sigma_{i}\}) =
(N_{HB}(\{\sigma_{i}\}),\,N_{h}(\{\sigma_{i}\}),\,N_{c}(\{\sigma_{i}\}))$,

 where $N_{HB}$, $N_{h}$, and $N_{c}$ denote the number
of hydrogen bonds, $h$-residues, and contacts, respectively, and the
corresponding control parameter vector ${\underline K} =
(\beta\,E_{HB},\,\, -\Delta
S(h),\,\, \beta\,k), $ with $\beta = (k_{B}T)^{-1}$, we obtain a
free energy of the  conformation $\{\sigma_{i}\}$,
\begin{equation}
\beta\,F(\{\sigma_{i}\}) = {\underline K} \cdot {\underline
S}(\{\sigma_{i}\}). \label{bF}
\end{equation}
The role of $F$ is  equivalent to that of a
Hamiltonian in other models.

This model has a coil-dominated high-temperature phase and a
helix-dominated
low-temperature phase.  For $k\ge0$ the low-temperature phase is a
single helix
spanning the whole system, and the change from coil to
helix is not accompanied by a phase transition\cite{LR61,PS70}.
For $k<0$, however, the low-temperature phase is multi-helical, with
neighboring helices stabilizing each other via tertiary interactions,
and the transition between
the coil and the multi-helical phases is of first order.

The multi-helical phase consists of several subphases characterized
by different numbers of helices.
These subphases are separated from each other, and from the coil
phase,
by significant barriers in $F$, a feature which leads to
long relaxation times at the transition point as well as to a slow,
glass-like, relaxation within the multi-helical phase\cite{EN}.
The existence of these subphases indicates that,
in addition to the order parameters that enter the Hamiltonian,
Eq.\ (\ref{bF}),
the number of helices of a conformation $\{\sigma_{i}\}$,
$N_{hel}(\{\sigma_{i}\})$, is an additional
relevant order parameter, see also Fig.\ \ref{hist}.

In order to investigate the behavior of this system at and around its
phase
transition point we generated order parameter histograms,
$n_{\underline K}({\underline S})$,
from MC simulations for parameter sets ${\underline K}$ corresponding
to the coil and
the multi-helical phase for various values of $L$.
For $L=40$ additional simulations close to the phase transition
point were performed for a check of the method. We
provide the respective parameters and histogram properties for that
length
in Table\ \ref{tab1}.

Using the MHM\cite{FS}, the relative probability to find the system
at
parameter set ${\underline K}$ in a state with order parameter
${\underline S}$ , $P_{{\underline K}}({\underline S})$,
can be approximated
from histograms obtained at the parameter sets ${\underline K}_i$,
$i=1,\ldots,M$, by
\begin{equation}
P_{{\underline K}}({\underline S}) =
{\sum_{i=1}^M g_i^{-1} n_i({\underline S}) \exp\left[-{\underline
K}\cdot {\underline S}\right] \over
 \sum_{j=1}^M g_j^{-1} N_j \exp\left[-{\underline K}_j\cdot
{\underline S} + f({\underline K}_j)\right] } .
\label{MHM}
\end{equation}
Here, $N_i=\sum_{{\underline S}}n_i({\underline S})$, $g_i=1+2\tau_i$
renormalizes the
size of the histogram to take correlation effects into account,
with $\tau_i$ being the correlation time of the simulation $i$. By
$f({\underline K})$
we denote the free energy of the system at parameter set ${\underline
K}$,
$\exp\left[-f({\underline K})\right]=Z({\underline
K})=\sum_{\underline S} P_{\underline K}({\underline S}),$
which is determined only up to an additive constant.

We have argued above that, using Eq.\ (\ref{MHM}), histograms of the
coil and the
multi-helical phase (histograms {\bf 1} and {\bf 3} of Tab.\
\ref{tab1})
suffice to reconstruct the first order phase transition. However,
this requires
the knowledge of the free energy difference between these histograms.
Applying
 Bennett's ARM\cite{Bennett76} to the model studied here leads to the
equation
\begin{equation}
\Delta f_{ij} =
\ln {{\left<{\cal F}({\underline S}\cdot\lbrack{\underline K}_{i}-
{\underline K}_{j}\rbrack+C)\right>_{i}}\over{ \left<{\cal
F}({\underline S}\cdot\lbrack
{\underline K}_{j}-{\underline K}_{i}\rbrack-C)\right>_{j}}}+C,
\label{ARM}
\end{equation}
where $\Delta f_{ij} = f({\underline K}_i) - f({\underline K}_j)$,
and
 $C=\ln(Z_{j}N_{i}/Z_{i}N_{j})$ has to be determined
self-consistently.
${\cal F}(x) = 1/\lbrack 1+\exp(x)\rbrack$ is the Fermi function, and
$\left<x\right>_{i}$ denotes  the average of $x$ with respect to
histogram $i$.
In order to apply Eq.\ (\ref{ARM}), the two histograms in
question have to overlap at least partially, which is not the case
for the
single-phase histograms {\bf 1} and {\bf 3}.
Bennett has already pointed out the possibility to obtain free energy
differences for disjoint histograms by performing additional
simulations
for intermediate parameter sets, so as to form an overlapping chain
of
histograms. However, a simulation near the critical point
is not feasible in general.

In our model we can exploit the  property that, by changing the
parameter
$\Delta S(h)$ to less negative values, systems are obtained which
exhibit
significantly shorter relaxation times and a more gradual
transition. For such systems, equilibrated histograms can be
generated for the separate phases {\it and} in the transition region
(see
Tab.\ \ref{tab1}) at a much smaller expense of computation time than
for the original
system.  Using such auxiliary histograms, a sequence of mutually
overlapping
histograms can be formed to join the single-phase histograms of the
original
system. This allows the determination of $\Delta f_{{\bf 31}}$
between
the disjoint histograms {\bf 1} and {\bf 3} of Tab.\ \ref{tab1} by
repeated application
of Eq.\ (\ref{ARM}), as illustrated in detail in Tab.\ \ref{tab2}.
For comparison, the value for
$\Delta f_{{\bf 31}}$ determined using histogram {\bf 2} of Tab.\
\ref{tab1} as
intermediate histogram is also given in Tab.\ \ref{tab2}. Both values
agree closely.
However, the expected error for the latter value is significantly
larger due to
the long relaxation time for the simulation near the critical point.

It was already noted that the multi-helical phase is also
characterized by
long relaxation times, here due to infrequent switches between
various
{\it coexisting} subphases
characterized by different helix numbers. These relaxation times
become prohibitively long particularly for systems with larger sizes
(we have
studied systems up to $L=200$).
With minor modifications, it is possible to apply the principles we
used to
reconstruct the histogram at the phase transition also for a
reconstruction of
histograms {\it within} the multi-helical phase. Histograms for
helical
 subphases with a constraint in helix number $N_{hel}$ can be
generated with
much smaller computational effort than for the full histogram, see
Tab.\ \ref{tab1}. They
are obtained simply by  introducing an infinite energy barrier for
all steps that attempt to change the helix number during the
simulations.

In direct analogy to the multiple-histogram equation, Eq.\
(\ref{MHM}), it is possible
to combine several subphase histograms $n_i^a({\underline S})$ with
identical helix
number $N_{hel}=a$, obtained at various parameter sets ${\underline
K}_i$, $i=1,\ldots, M$.
The relative probability {\it within
this subphase} of a system state ${\underline S}$ (with $N_{hel}=a$),
at
 parameter set ${\underline K}$, is approximated by
\begin{equation}
P_{{\underline K}}^{a}({\underline S}) =
{\sum_{i=1}^M (g_i^a)^{-1} n_i^a({\underline S})
\exp\left[-{\underline K}\cdot {\underline S}\right] \over
 \sum_{j=1}^M (g_j^a)^{-1} N_j^a \exp\left[-{\underline K}_j\cdot
{\underline S} + f^a({\underline K}_j)\right] } .
\label{sub}
\end{equation}
\noindent Here, the sum in the denominator ranges only over subphase
histograms with $N_{hel}=a$. We note that the free energies
 $f^a({\underline K}_j)$ in Eq.\ (\ref{sub}) have now to be
calculated
with respect to subphase histograms with $N_{hel} = a$ only.

These subphase histograms can now be combined, again using ideas of
the MHM.
Following the arguments in Ref.\ \cite{FS}, and using the property
that the subphases
fully partition the order parameter space into {\it disjoint}
patches, one
arrives at
\begin{equation}
P_{{\underline K}}({\underline S}) = \sum_{a} p_{{\underline K}}^a \,
P_{{\underline K}}^a({\underline S}) \label{combine}
\end{equation}
for the combined histogram. Given a full histogram of an auxiliary
system at
some convenient parameter value ${\underline K}'$, one obtains for
the relative
contribution of subphase $a$, $p_{{\underline K}}^a$ in Eq.\
(\ref{combine}),
\begin{equation}
 p_{{\underline K}}^a = p_{{\underline K}'}^a \exp \lbrack
f^a({\underline K}) - f^a({\underline K}')\rbrack \,,\label{combine2}
\end{equation}
where $p_{{\underline K}'}^a$ is given by $p_{{\underline K}'}^a =
\sum_{{\underline S}}n_{{\underline K}'}^a({\underline S}) /
\sum_{{\underline S}}n_{{\underline K}'}({\underline S})$.

Using the methods described we have investigated the first order
transition of our model. Contrary to other
approaches\cite{Binder}, we choose the fluctuations of the specific
energy
$e=(N_{HB} \cdot E_{HB}+N_c \cdot k)/L$,
{\it i.e.} $e_{2}=\left<(e-\left<e\right>)^2\right>$, as the
observable to
monitor the transition\cite{Hueller94}. For first order transitions
the
peak of this function assumes a finite nonzero value in the
thermodynamic
limit, which is given by $e_{2,max}=(\Delta e)^2/4$, where
$\Delta e$ is the specific latent heat. Figure\ \ref{fig1} shows the
behavior of $e_2$ for
various values of the system size.
For $L=40$ the curves deriving from histogram {\bf2}, from the
combination of
histograms {\bf1} and {\bf3}, and from {\bf1} together with {\bf3.2}
to {\bf
3.5}, respectively, all coincide, thereby confirming our approach.
For the longer systems subphase histograms for the helical phase were
used
invariably. Included in the figure is
the extrapolated function for infinite system size, corresponding to
a latent
heat of $\Delta e =1.37\,E_{HB}$ and a transition at
$\beta=2.78\,E_{HB}$.

In closing, we will briefly discuss why other methods proposed for
obtaining simulation results for first-order phase transitions
fail in our system. Due to the high
dimensionality of the order parameter space and the complicated phase
structure, compare Fig.\ \ref{hist}, the straightforward
identification of a --- possibly
small number --- of transition states in order parameter space,
necessary
for a successful application of the multicanonical method\cite{BN},
is
impossible. In addition, the states corresponding to the individual
phases
occupy only an extremely small part of the order parameter space. By
raising the
temperature, as in the entropic
sampling\cite{Lee93} and simulated tempering\cite{MP92} methods, not
only the
transition states, but also an enormous number of ``uninteresting"
states would
become accessible. To suppress
these states at higher temperatures, a prohibitively large number
of parameters (of the order of $10^6$ for $L=100$) would have to be
optimized.
The microcanonical ensemble approach promoted by H\"uller and
coworkers\cite{Hueller94,Hueller} or cluster dynamics
approaches\cite{KW93,SW}
are also not applicable in our case.

We believe that the approach presented here offers new possibilities
for
investigating model systems in parameter regimes where relaxation
times are
too long to allow equilibrated MC simulations,
like at strong first-order transitions or at glass transitions
and within the glass phase. A prerequisite for the successful
application
of the method is the existence of a detour in control parameter
space with a more gradual transition, {\it i.e.}, much shorter
relaxation
times, so
that the necessary free energy differences can be determined.
In order to apply the subphase method, Eqs.\ (\ref{sub}) to
(\ref{combine2}), there has to be
an order parameter set which allows to identify subphases
unambiguously.

M. E. gratefully acknowledges  support by a stipend from the
Studienstiftung
des  Deutschen Volkes. The authors thank K. Neymeyr for providing
additional computational
resources.

\begin{figure}
\caption{Reconstructed histogram for $L=40$, $\beta \vert E_{HB}\vert
= 3.4$ (transition region), projected onto the $(F, N_{hel})$ plane.}
\label{hist}
\end{figure}

\begin{figure}
\caption{Fluctuations in specific energy,
$e_2=\left<(e-\left<e\right>)^2\right>$, $vs$
$\beta$ for $L$ as indicated; also shown is the extrapolation for
$L\to\infty$.}
\label{fig1}
\end{figure}

\begin{table}
\caption{Histograms, $L=40$.}
\label{tab1}
\begin{tabular}{rdccr}
& $\beta$ &  Size & Phase  & Correlation time  \\
& $\lbrack \vert E_{HB}\vert^{-1}\rbrack$ & {$\lbrack10^6$MC
steps$\rbrack$} &   & {$\lbrack$MC steps$\rbrack$} \\
\tableline
& & & & \\
 \multicolumn {5}{c}{Histograms, $\Delta S(h)/k_B = -3.57$} \\
& & & & \\
 {\bf 1} & 2.66 & 10 & random coil  & 30  \\
{\bf  2} & 3.43  & 600 & mixed  & 350000  \\
{\bf  3} & 3.96  & 50 & multi-helical & 4600  \\
   & & & &\\
 \multicolumn {5}{c}{Auxiliary histograms, $\Delta S(h)/k_B = -0.57$}
 \\
  & & & &\\
{\bf  1a} & 0.16 &  10 & random coil &  50  \\
{\bf  2a} & 1.16  & 50 & mixed  & 200  \\
{\bf  3a} & 1.56  & 20 & multi-helical & 300  \\
& & & & \\
\multicolumn {5}{c}{Subphase histograms, $\Delta S(h)/k_B = -3.57$}
\\
& & & & \\
{\bf  3.2} & 3.96 &5 & $N_{hel}=2$  & 100  \\
{\bf  3.3} & 3.96 & 5 & $N_{hel}=3$  & 100  \\
{\bf  3.4} &  3.96 & 5 & $N_{hel}=4$  & 160  \\
{\bf  3.5} &  3.96 & 5 & $N_{hel}=5$ & 170  \\
\end{tabular}
\end{table}

\begin{table}
\caption{$\Delta f$, $L=40$, according to Eq.\ (\protect\ref{ARM}).
\protect\\
Root mean square deviation (rmsd) according to Bennett
\protect\cite{Bennett76}.}
\label{tab2}
\begin{tabular}{rdr}
 Histograms & $\Delta f$ & rmsd \\
\tableline
\\
{\bf 1/2} &   -1.627 & $7 \cdot 10^{-2}$ \\
 {\bf 2/3} &   -22.884 & $3 \cdot 10^{-2}$ \\
 $\Sigma$:{\bf 1/3} &   -24.511 & $7.5 \cdot 10^{-2}$ \\
  & & \\
 {\bf 1/1a} &   -9.501  & $1.3 \cdot 10^{-2}$\\
 {\bf 1a/2a} &   -3.828  &   $8 \cdot 10^{-3}$ \\
 {\bf 2a/3a} &   -14.741  & $5.2 \cdot 10^{-3}$\\
 {\bf 3a/3} &   3.569 &  $9.3 \cdot 10^{-3}$ \\
 $\Sigma$:{\bf 1/3} &   -24.501 & $1.9 \cdot 10^{-2}$ \\
\end{tabular}
\end{table}

\end{document}